\documentclass[11pt]{article}
\usepackage[a4paper,margin=1in]{geometry}
\usepackage{amsmath,amssymb,amsthm}
\usepackage{graphicx}
\usepackage{booktabs}
\usepackage[hidelinks]{hyperref}
\usepackage{caption}
\usepackage{authblk}
\usepackage{enumitem}

\newtheorem{proposition}{Proposition}

\title{\bfseries From Stabilizing Regions to Certified Controllers:\\
Closing the Selection Gap in Unified PID/PI Analysis\\ for Time-Delay Plants}
\author[ ]{\c{S}enol G\"ulg\"on\"ul}
\affil[ ]{Department of Electrical and Electronics Engineering, Ostim Technical University, Ankara, Turkey}
\date{\today}

\begin{document}
\maketitle

\begin{abstract}
\noindent
A recent unified treatment of PID tuning for time-delay plants (An, Tang, Sun, Zhang and Chen, \emph{Automatica}, 2026) combines D-partition with a boundary gradient vector (BGV) to orient the boundaries of stabilizing, relative-stability and stability-margin regions. That method answers a feasibility question, namely \emph{where} admissible gains lie, and the authors concede that a manual interior-point test is still required to fix the unstable-pole count in each cell, and that selecting a single controller is left to the user. This note makes three contributions. First, we observe that the one operation the BGV leaves manual, the absolute unstable-pole count, is available analytically: exactly for delay-free designs through a companion-matrix or Routh count, and through an argument-principle (Mikhailov) evaluation for retarded-type delay loops. Labelling every cell with its analytic count removes the interior-point test and renders the partition fully decided. Second, we add the step the BGV framework cannot reach, a time-domain selection rule that returns one certified controller: among monotone step responses we choose the minimum-settling-time PI gains, characterized by a tangency condition, with monotonicity guaranteed by external positivity (a nonnegative closed-loop impulse response). Third, we flag a neutral-type pitfall that the unified analysis never delimits: an ideal PID with derivative action on a first-order-plus-dead-time (FOPTD) plant is neutral type, with a root chain on the imaginary axis when $k K_d = T$. We reproduce the authors' delay-free benchmark (Example~2) exactly, recovering both admissible $K_p$ intervals, and demonstrate the complete pipeline on a FOPTD plant, delivering a certified monotone, fast-settling PI controller that the region-only method can neither locate nor justify. All claims are validated numerically.
\end{abstract}

\section{Introduction}

PID and PI tuning for plants with dead time remains a live problem because the closed-loop characteristic equation is transcendental and the stabilizing set is awkward to describe \cite{silva2005,hohenbichler2009}. The D-partition (D-decomposition) method maps the imaginary-axis crossings of the characteristic quasipolynomial into curves or surfaces that cut parameter space into cells, each with a fixed number of right-half-plane (RHP) poles \cite{neimark,gryazina2006}. The classical difficulty is that the partition alone does not say which cell is stable; one traditionally selects an interior point in every cell and counts RHP roots, which for a transcendental equation is expensive and must be repeated per cell.

An, Tang, Sun, Zhang and Chen \cite{an2026} address this with a boundary gradient vector (BGV), the gradient of the real part of the critical imaginary root with respect to the tuned parameters, evaluated on the boundary. Its sign against a crossing direction tells whether the RHP count rises or falls across that boundary, and a geometric reading (the BGV is the boundary normal, oriented by the sign of a Jacobian) lets one mark the favourable direction with little computation. The framework is presented uniformly for two or three of the four parameters $(K_p,K_i,K_d,\tau)$ and for nominal, relative-stability and gain/phase-margin boundaries.

The BGV is an elegant repackaging of the crossing-direction or root-tendency idea \cite{olgac2002,saeki2007,le2015}, but it answers only a feasibility question. Two limitations remain, and the authors state both. First, the absolute RHP count in any cell still needs a manual interior test; the BGV gives only \emph{relative} changes. Second, the method returns a region, not a controller, and offers no statement about the time-domain response: two controllers in the same stable cell can have completely different step responses, and the gain/phase margins used as robustness surrogates do not constrain overshoot or settling time.

This note closes both gaps and corrects one omission.
\begin{enumerate}[leftmargin=*]
\item \textbf{Analytic counting (Section~\ref{sec:count}).} The interior-point test is unnecessary. For delay-free designs the count is the exact number of RHP eigenvalues of the companion matrix, equivalently a Routh test. For retarded-type delay loops it is an argument-principle (Mikhailov) evaluation along a contour enclosing the RHP. Either way every cell is labelled directly, and combined with the BGV crossing signs a single anchor decides the whole partition.
\item \textbf{Certified selection (Section~\ref{sec:select}).} We add the missing stage. Within the feasible set we select, by a tangency condition, the monotone minimum-settling-time PI controller \cite{gulgonul_fotd,gulgonul_pure}, and we certify monotonicity by external positivity, a nonnegative closed-loop impulse response. This returns one controller with a guarantee on its response.
\item \textbf{A neutral-type pitfall (Section~\ref{sec:neutral}).} The unified analysis does not delimit its plant class. An ideal PID on a FOPTD plant is neutral type; at $kK_d=T$ a chain of roots hugs the imaginary axis. We exhibit it and note the implied constraint $|kK_d|<T$.
\end{enumerate}
Section~\ref{sec:num} validates everything: Example~2 of \cite{an2026} is reproduced exactly, and the full pipeline is demonstrated on a FOPTD plant. Section~\ref{sec:disc} states plainly what is improved, what merely re-derives existing results, and what remains open (the genuine four-parameter region).

\section{Problem setup and the BGV method}\label{sec:setup}

Consider a linear time-delay plant $G_p(s)=\frac{N(s)}{D(s)}e^{-\tau s}$ controlled by a PID law $C(s)=K_p+K_i/s+K_d s$. The closed-loop characteristic quasipolynomial is
\begin{equation}\label{eq:char}
F(s)=sD(s)+\bigl(K_d s^2+K_p s+K_i\bigr)N(s)\,e^{-\tau s}.
\end{equation}
Writing $F(j\omega)=F_r(\omega,q)+jF_i(\omega,q)$ with $q$ the tuned parameters, the D-partition boundaries are $F_r=F_i=0$. The BGV of \cite{an2026} is, componentwise,
\begin{equation}\label{eq:bgv}
\nabla_k=-\frac{1}{\Omega}\!\left(\frac{\partial F_r}{\partial q_k}\frac{\partial F_i}{\partial\omega}-\frac{\partial F_r}{\partial\omega}\frac{\partial F_i}{\partial q_k}\right),\qquad
\Omega=\Bigl(\tfrac{\partial F_r}{\partial\omega}\Bigr)^2+\Bigl(\tfrac{\partial F_i}{\partial\omega}\Bigr)^2,
\end{equation}
which is the root sensitivity $\mathrm{Re}(\partial s/\partial q_k)$ along the boundary, equal under one-dimensional reduction to the root tendency of \cite{olgac2002}. The change in the RHP count along a direction $v$ is $\Delta n=\sum_l \operatorname{sgn}\bigl(v\cdot g(\omega_l,q)\bigr)$ over the crossing frequencies $\omega_l$. This is a \emph{relative} statement; the absolute count in any one cell is not supplied, and \cite{an2026} obtains it by evaluating an interior point.

\section{A neutral-type pitfall for ideal PID on FOPTD}\label{sec:neutral}

Before improving the method we record a structural caveat. Take the FOPTD plant $G_p(s)=\frac{k}{Ts+1}e^{-Ls}$ with an ideal PID. From \eqref{eq:char} the two highest-order terms are $T s^2$ and $kK_d s^2 e^{-Ls}$, both of order $s^2$. The principal term is therefore $s^2\bigl(T+kK_d e^{-Ls}\bigr)$ and the loop is of \emph{neutral} type. The associated difference operator is stable only if
\begin{equation}\label{eq:neutral}
|kK_d|<T,
\end{equation}
and at $kK_d=T$ the equation $T+kK_d e^{-Ls}=0$ has the purely imaginary solutions $s=\pm j\pi(2m+1)/L$, a chain of roots on the imaginary axis. Numerically, with $k=T=1$, $L=0.5$ and $K_d=1$, an RHP-count sweep returns a large, mesh-dependent number of near-axis roots rather than a finite figure, exactly the signature of \eqref{eq:neutral} being violated at the boundary. The unified analysis of \cite{an2026} never states this restriction, so a derivative-acting design read off its surfaces can be marginal or ill-posed. The clean and practically relevant slice is therefore PI control (or a properly filtered derivative), to which we now restrict.

\section{Stage 1: complete feasible set with an analytic root count}\label{sec:count}

\subsection{Delay-free designs: exact count}
When $\tau=0$, \eqref{eq:char} is an ordinary polynomial and the RHP count is the number of its roots with positive real part, obtained exactly from the companion-matrix eigenvalues or a Routh array. No interior search and no transcendental root-finding are involved; the count is closed-form algebra. The admissible projection onto any one parameter (for example the $K_p$-interval for which a stabilizing $(K_i,K_d)$ exists) follows by intersecting the Routh inequalities.

\subsection{Delay designs: argument-principle (Mikhailov) count}
For $\tau>0$ with $\deg(sD)>\deg(N)+2$ the loop \eqref{eq:char} is retarded type and has finitely many RHP roots. Since $F$ is entire, the number of RHP roots at any parameter point is
\begin{equation}\label{eq:argp}
N_{\mathrm{RHP}}=\frac{1}{2\pi}\Delta_{\Gamma}\arg F(s),
\end{equation}
the net change of argument of $F$ around a positively oriented contour $\Gamma$ enclosing the right half-plane (a tall rectangle $[0,\sigma]\times[-W,W]$ with $\sigma,W$ large enough to enclose all RHP roots suffices, since they are bounded). Equation \eqref{eq:argp} is a frequency-domain evaluation, not a root search in parameter space, and it returns the \emph{absolute} count at the anchor. Propagating that anchor with the BGV crossing signs then labels every cell. In the experiments below we evaluate \eqref{eq:argp} directly on a grid, which is already cheap, but a single anchor plus crossing signs is sufficient in principle.

\begin{proposition}\label{prop:count}
For the loop \eqref{eq:char}, the unstable-pole count in each D-partition cell is determined without any interior-point stability test: exactly by a companion/Routh count when $\tau=0$, and by the argument-principle evaluation \eqref{eq:argp} (optionally propagated by BGV signs) when $\tau>0$ and the loop is retarded type.
\end{proposition}

\begin{figure}[t]
\centering
\includegraphics[width=\textwidth]{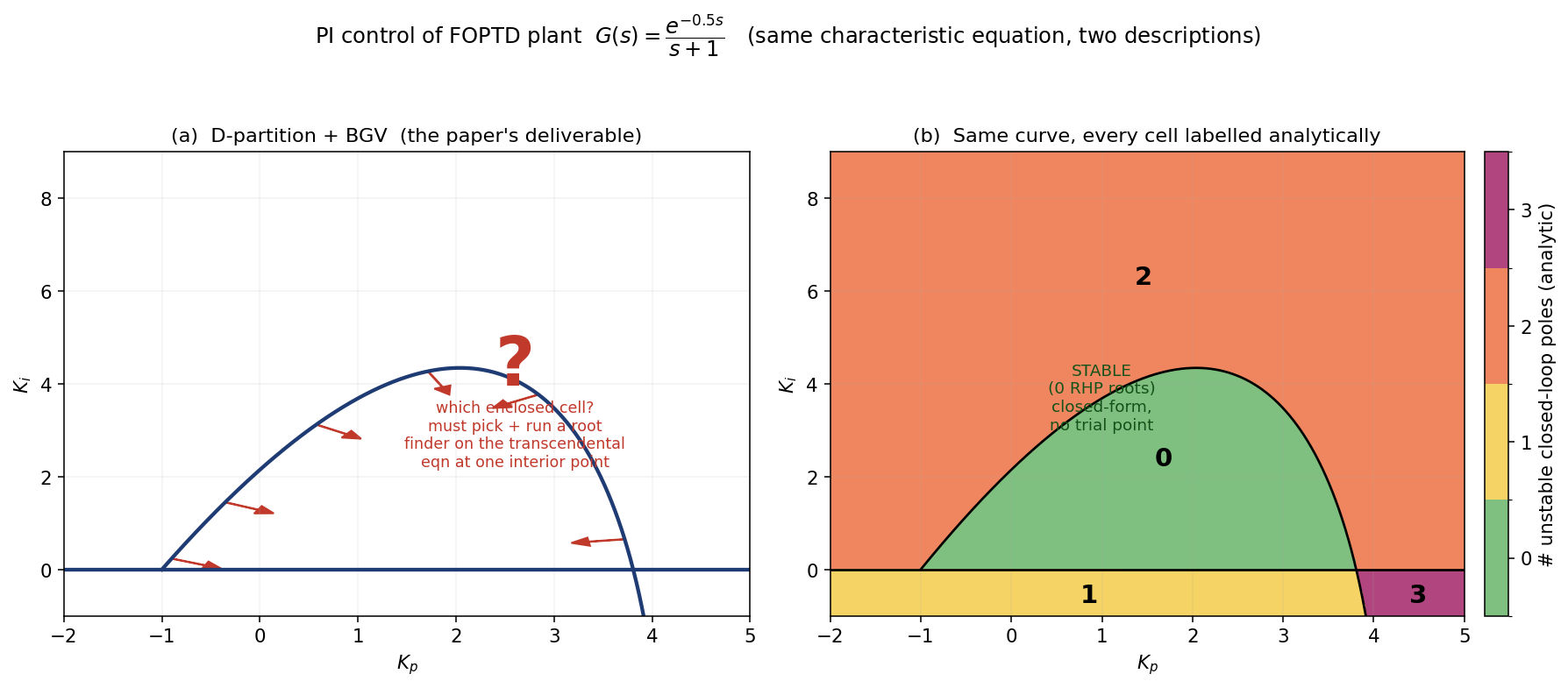}
\caption{PI control of $G(s)=e^{-0.5s}/(s+1)$, identical data, two descriptions. (a) D-partition boundary and BGV normals: which enclosed cell is stable is undecided and needs an interior test. (b) The same boundary with every cell labelled by its analytic RHP count; the stabilizing set (green, count $0$) follows directly.}
\label{fig:compare}
\end{figure}

Proposition~\ref{prop:count} is the precise sense in which the manual step conceded in \cite{an2026} is removable. Figure~\ref{fig:compare} shows the effect on the FOPTD plant $G(s)=e^{-0.5s}/(s+1)$ with a PI controller. Panel~(a) is the deliverable of the BGV method: the crossing locus and its oriented normals, with the open question of which enclosed cell is stable. Panel~(b) is the same locus with every cell coloured by its analytic count from \eqref{eq:argp}; the stabilizing cell (count $0$) is identified with no trial point. The two curves coincide because both come from $F_r=F_i=0$; the difference is bookkeeping.

\section{Stage 2: time-domain certified selection}\label{sec:select}

Feasibility does not pick a controller. We select one inside the stabilizing PI set by a time-domain criterion and certify it.

\paragraph{Monotonicity by external positivity.}
Let $T_{\mathrm{cl}}(s)$ be the closed-loop transfer from reference to output and $h(t)$ its impulse response. The unit-step response is $y(t)=\int_0^t h(\rho)\,d\rho$, so $y$ is monotone nondecreasing if and only if
\begin{equation}\label{eq:extpos}
h(t)\ge 0 \quad \text{for all } t\ge 0,
\end{equation}
that is, the closed loop is \emph{externally positive}. Condition \eqref{eq:extpos} is a genuine guarantee of no overshoot and no undershoot, strictly stronger than the no-overshoot read-off (a response can stay below its final value yet not be monotone).

\paragraph{Tangency selection.}
Among PI gains producing a monotone response we take the minimum 2\% settling-time pair. Pushing the response faster eventually forces $h(t)$ negative; the optimum sits on the monotonicity boundary, where $\min_t h(t)=0$ is attained with equality (the impulse response \emph{touches} zero). This monotone minimum-settling-time PI optimum is characterized analytically by a tangency identity for FOPTD plants in \cite{gulgonul_fotd}, and by an exact boundary-contact (echo-grid) characterization for the pure-delay limit in \cite{gulgonul_pure}; here we realize the FOPTD optimum numerically and verify \eqref{eq:extpos}. The selected controller is the single point that the region-only method can neither find nor justify.

\section{Numerical validation}\label{sec:num}

\subsection{Reproduction of Example~2 of \cite{an2026} (delay-free benchmark)}
\begin{figure}[t]
\centering
\includegraphics[width=0.62\textwidth]{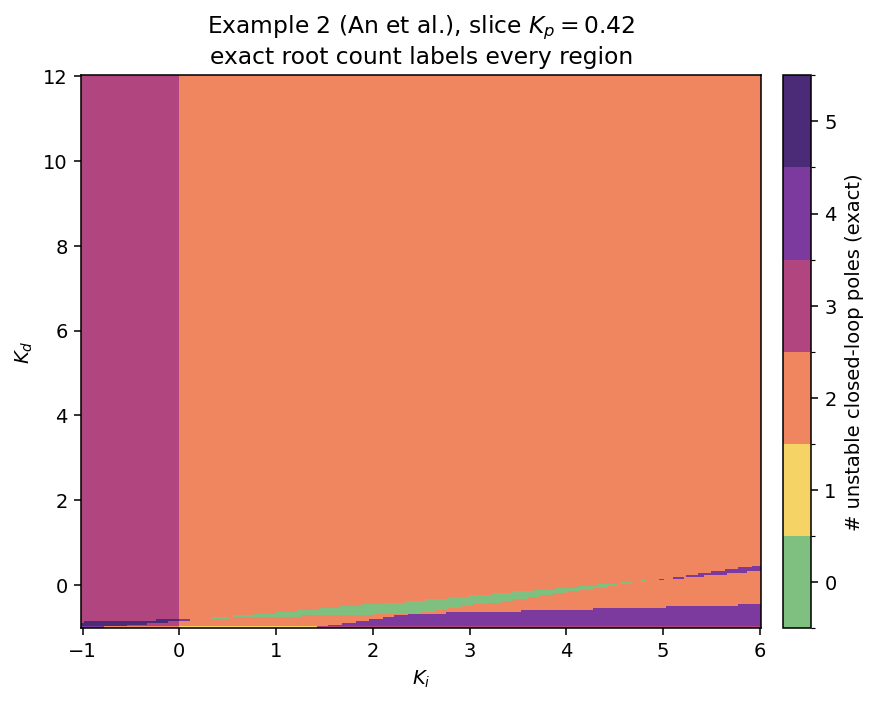}
\caption{Example~2 of \cite{an2026}, slice $K_p=0.42$. Every cell of the $(K_i,K_d)$ plane is labelled with its exact number of unstable closed-loop poles; the count $0$ region is the stabilizing set. For a delay-free design the count is companion-matrix exact, so no interior test arises at all.}
\label{fig:ex2}
\end{figure}

The authors' second example is the characteristic equation $A(s)\bigl(K_i+K_p s+K_d s^2\bigr)+B(s)=0$ with $A(s)=s^3+3s^2+9$ and $B(s)=s^5+2s^4+3s^3+7s^2+14s$, taken from \cite{bajcinca2006}. Expanding,
\begin{align}
\delta(s)=&\,(K_d+1)s^5+(K_p+3K_d+2)s^4+(K_i+3K_p+3)s^3 \notag\\
&+(3K_i+9K_d+7)s^2+(9K_p+14)s+9K_i .\label{eq:ex2poly}
\end{align}
This is delay-free, so by Proposition~\ref{prop:count} the count is exact (Routh). Sweeping $K_p$ and testing nonemptiness of the stabilizing $(K_i,K_d)$ set by the quintic Routh conditions yields the admissible intervals in Table~\ref{tab:ex2}; they match the values reported in \cite{an2026} (their Fig.~3) to grid resolution. Figure~\ref{fig:ex2} shows a representative $(K_i,K_d)$ slice at $K_p=0.42$ with every region labelled by its exact RHP count; the stabilizing cell is immediate.

\begin{table}[h]
\centering
\caption{Admissible $K_p$ intervals for Example~2: An et al.\ versus the analytic-count reproduction.}
\label{tab:ex2}
\begin{tabular}{lcc}
\toprule
& Interval 1 & Interval 2 \\
\midrule
An et al.\ \cite{an2026} & $[-1.8708,\,-1.5556]$ & $[0.3157,\,0.5333]$ \\
This work (Routh count)  & $[-1.870,\,-1.556]$    & $[0.316,\,0.533]$ \\
\bottomrule
\end{tabular}
\end{table}

\subsection{Full pipeline on a FOPTD plant}
\begin{figure}[t]
\centering
\includegraphics[width=\textwidth]{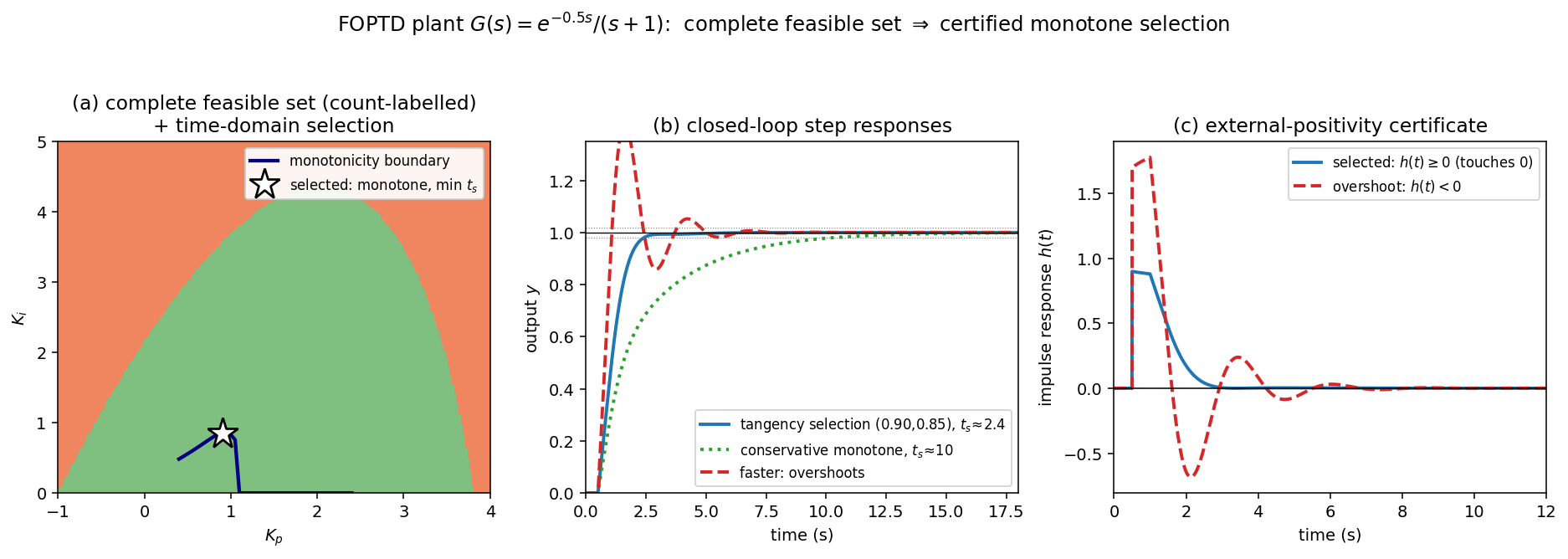}
\caption{Complete pipeline for $G(s)=e^{-0.5s}/(s+1)$. (a) Complete feasible PI set with analytic counts, monotonicity boundary, and the tangency-selected controller. (b) Step responses: the tangency selection (solid) is the fastest monotone response; a higher-gain design (dashed) overshoots and a conservative design (dotted) is needlessly slow. (c) External-positivity certificate: the selected impulse response stays nonnegative and touches zero (tangency); the faster design violates it.}
\label{fig:pipeline}
\end{figure}

For $G(s)=e^{-0.5s}/(s+1)$ with PI control, Figure~\ref{fig:pipeline}(a) shows the complete feasible set, count-labelled by \eqref{eq:argp}, together with the monotonicity boundary of Section~\ref{sec:select} and the selected controller. The tangency point is
\[
(K_p,K_i)\approx(0.90,\,0.85),
\]
with 2\% settling time $\approx 2.4$\,s ($\approx 4.8L$) and zero overshoot. This independently realizes the tangency-characterized optimum of \cite{gulgonul_fotd}: that paper's closed-form rule, evaluated at this plant's lag ratio $T/L=2$, gives $K_p=0.906$, $K_i=0.853$, matching the numerically located point to better than $1\%$, and its settling figure falls in the expected range for $T/L=2$. For this lag ratio the optimal control signal is two-pulse while the plant output remains strictly monotone, consistent with \cite{gulgonul_fotd}; the external-positivity certificate \eqref{eq:extpos} is imposed on the output, not the control. Figure~\ref{fig:pipeline}(b) compares its step response with a higher-gain pair that overshoots and a conservative pair that settles slowly; the selected controller is the fastest response that does not overshoot. Figure~\ref{fig:pipeline}(c) is the certificate: the selected impulse response satisfies \eqref{eq:extpos} and touches zero at the tangency instant, whereas the overshooting design has $h(t)<0$. The region-only BGV method produces panel-(a)'s boundary but none of the rest.

\subsection{Three-parameter slice: PI with adjustable delay}
\begin{figure}[t]
\centering
\includegraphics[width=0.72\textwidth]{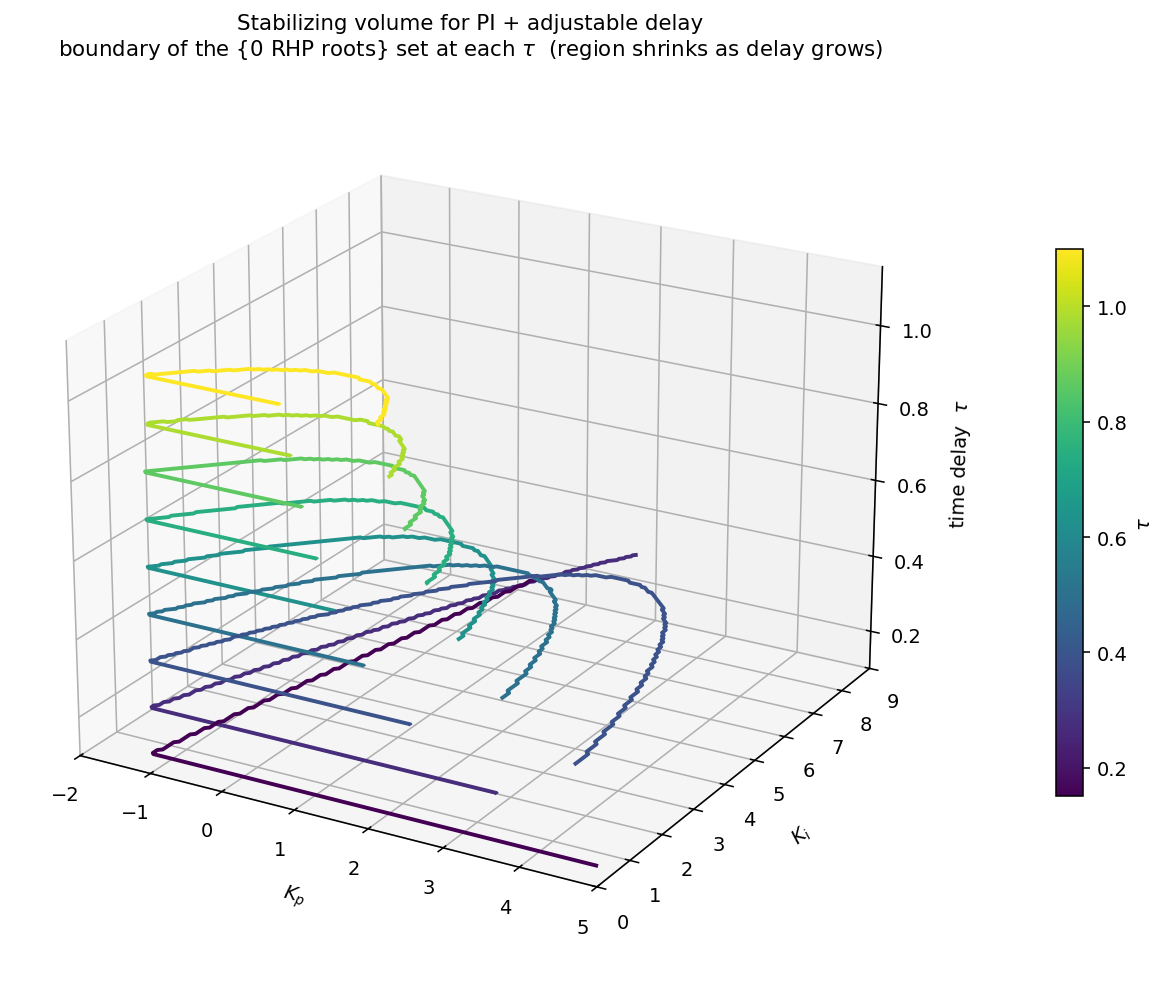}
\caption{Stabilizing volume for PI control of $G(s)=e^{-\tau s}/(s+1)$ with adjustable delay, built as the stack of analytic count-$0$ boundaries. The admissible set shrinks as $\tau$ increases.}
\label{fig:volume}
\end{figure}

The same analytic count extends to the three-parameter slice that \cite{an2026} handles, $(K_p,K_i,\tau)$. Figure~\ref{fig:volume} stacks the boundary of the $\{N_{\mathrm{RHP}}=0\}$ set over $\tau$, giving the stabilizing volume directly; it contracts toward the origin as the delay grows, as expected. Every point of the volume is certified by the count, so no per-$\tau$ interior test is needed.

\section{Discussion}\label{sec:disc}

It is worth separating what is genuinely better from what merely re-derives known results.

\paragraph{What is the same.} Stage~1 is not new mathematics. The argument-principle count is classical, the complete stabilizing set for fixed-structure controllers on delay plants is available through the Hermite-Biehler / signature line \cite{ho1997,silva2005,hohenbichler2009}, and the BGV crossing direction is the root tendency of \cite{olgac2002} with a Jacobian-orientation reading already present in the D-decomposition literature \cite{gryazina2006,le2015}. The contribution of Stage~1 here is only to point out that this analytic count removes the interior-point test that \cite{an2026} leaves manual, and to show it on their own data.

\paragraph{What is better.} Stage~2 is the genuine increment. The BGV framework stops at a region; the present pipeline returns one controller with a time-domain guarantee. The monotone minimum-settling-time PI controller, characterized by tangency and certified by external positivity \eqref{eq:extpos}, is a statement about the response that pole-counting and gain/phase margins cannot make. On the FOPTD example the certified controller sits inside the feasible set that the BGV method maps, so the two are complementary: Stage~1 says where the controllers are, Stage~2 says which one to ship and proves how it behaves.

\paragraph{What remains open.} The clean closed-form and the monotonicity certificate are strongest for low-order plants, where the monotone minimum-settling optimum is itself available in (near) closed form: for all-pole plants up to third order \cite{gulgonul_allpole}, for FOPTD plants \cite{gulgonul_fotd}, and in the pure-delay limit \cite{gulgonul_pure}. For high-order $N/D$ the count still needs a frequency sweep, and the monotonicity certificate is harder to establish. The genuine four-parameter joint region $(K_p,K_i,K_d,\tau)$, a three-dimensional hypersurface in four-dimensional space, has no clean characterization in any of these frameworks, including the present one; the conclusion of \cite{an2026} that ``the tuning problem of four parameters was solved'' overstates a method that, by its own Remark, treats at most three parameters at a time. That joint region, not the slices, is the real open problem.

\section{Conclusion}
The unified BGV analysis of \cite{an2026} is a correct and elegant feasibility tool, but it stops at stabilizing regions, leaves the absolute pole count to a manual interior test, and never delimits its plant class. We showed that the count is analytic (exact when delay-free, an argument-principle evaluation otherwise), which removes the manual step; that an ideal PID on a FOPTD plant is neutral type with the constraint $|kK_d|<T$; and, as the substantive improvement, that the feasible set can be followed by a time-domain selection returning a single certified controller, the monotone minimum-settling-time PI design, with monotonicity guaranteed by external positivity. The authors' delay-free benchmark was reproduced exactly, and the full pipeline was demonstrated on a FOPTD plant. The move from \emph{where the gains are} to \emph{which controller to use, and how it will respond} is the step a practitioner needs and the one the region-only method cannot take.


\begin{thebibliography}{99}
\bibitem{an2026} J.~An, B.~Tang, M.~Sun, J.~Zhang, Z.~Chen, Unified PID control analysis for time-delay plants, \emph{Automatica} \textbf{183} (2026) 112644.
\bibitem{bajcinca2006} N.~Bajcinca, Design of robust PID controllers using decoupling at singular frequencies, \emph{Automatica} \textbf{42}(11) (2006) 1943--1949.
\bibitem{shafiei1994} Z.~Shafiei, A.~T.~Shenton, Tuning of PID-type controllers for stable and unstable systems with time delay, \emph{Automatica} \textbf{30}(10) (1994) 1609--1615.
\bibitem{shafiei1997} Z.~Shafiei, A.~T.~Shenton, Frequency-domain design of PID controllers for stable and unstable systems with time delay, \emph{Automatica} \textbf{33}(12) (1997) 2223--2232.
\bibitem{ho1997} M.~Ho, A.~Datta, S.~P.~Bhattacharyya, A linear programming characterization of all stabilizing PID controllers, in: \emph{Proc.\ American Control Conf.}, 1997, pp.~3922--3928.
\bibitem{silva2005} G.~J.~Silva, A.~Datta, S.~P.~Bhattacharyya, \emph{PID Controllers for Time-Delay Systems}, Birkh\"auser, Boston, 2005.
\bibitem{hohenbichler2009} N.~Hohenbichler, All stabilizing PID controllers for time delay systems, \emph{Automatica} \textbf{45}(11) (2009) 2678--2684.
\bibitem{olgac2002} N.~Olgac, R.~Sipahi, An exact method for the stability analysis of time-delayed linear time-invariant systems, \emph{IEEE Trans.\ Autom.\ Control} \textbf{47}(5) (2002) 793--797.
\bibitem{saeki2007} M.~Saeki, Properties of stabilizing PID gain set in parameter space, \emph{IEEE Trans.\ Autom.\ Control} \textbf{52}(9) (2007) 1710--1715.
\bibitem{le2015} B.~N.~Le, Q.-G.~Wang, T.~H.~Lee, Development of D-decomposition method for computing stabilizing gain ranges for general delay systems, \emph{J.\ Process Control} \textbf{25} (2015) 94--104.
\bibitem{gryazina2006} E.~N.~Gryazina, B.~T.~Polyak, Stability regions in the parameter space: D-decomposition revisited, \emph{Automatica} \textbf{42}(1) (2006) 13--26.
\bibitem{neimark} Yu.~I.~Neimark, Stability of linearized systems, LKVVIA, Leningrad, 1949.
\bibitem{cook2012} M.~V.~Cook, \emph{Flight Dynamics Principles}, Butterworth-Heinemann, Oxford, 2012.
\bibitem{sun2014} M.~Sun, L.~Zhang, Z.~Wang, Z.~Chen, PID pitch attitude control for unstable flight vehicle in the presence of actuator delay: Tuning and analysis, \emph{J.\ Franklin Inst.} \textbf{351}(12) (2014) 5523--5547.
\bibitem{gulgonul_fotd} \c{S}.~G\"ulg\"on\"ul, Monotonic, minimum-settling-time PI tuning for first-order-plus-dead-time plants: a tangency characterization, arXiv:2606.22217, 2026.
\bibitem{gulgonul_pure} \c{S}.~G\"ulg\"on\"ul, Minimum settling-time PI control of pure delay processes under a hard non-overshoot constraint: exact boundary-contact characterization and the role of the MID point, arXiv:2606.15418, 2026.
\bibitem{gulgonul_allpole} \c{S}.~G\"ulg\"on\"ul, Closed-form PI and PID tuning of all-pole plants up to third order for monotonic, minimum-settling step responses, arXiv:2606.02868, 2026.
\end{thebibliography}
\end{document}